\documentclass[aps,prd,preprint,superscriptaddress,tightenlines,nofootinbib]{revtex4}
\usepackage{epsfig}
\usepackage{graphicx}
\usepackage{dcolumn}
\usepackage{bm}

\newcommand{\etal}{{\it et al.}}

\begin{document}


\preprint{\tighten\vbox{\hbox{\hfil CLNS 06/1975}
                        \hbox{\hfil CLEO 06-17}}}


\title
{\LARGE Measurement of Inclusive Production of
\boldmath{$\eta$,~$\eta'$ and $\phi$} Mesons in \boldmath{$D^0$,
$D^+$} and $D_s^+$ Decays}\


\author{G.~S.~Huang}
\author{D.~H.~Miller}
\author{V.~Pavlunin}
\author{B.~Sanghi}
\author{I.~P.~J.~Shipsey}
\author{B.~Xin}
\affiliation{Purdue University, West Lafayette, Indiana 47907}
\author{G.~S.~Adams}
\author{M.~Anderson}
\author{J.~P.~Cummings}
\author{I.~Danko}
\author{J.~Napolitano}
\affiliation{Rensselaer Polytechnic Institute, Troy, New York 12180}
% ...
\collaboration{CLEO Collaboration} 
\noaffiliation


\date{October 4, 2006}

\begin{abstract}
We measure the inclusive branching fractions of charm mesons into
three mesons with large $s\overline{s}$ content, namely the $\eta$,
$\eta'$ and $\phi$. Data were accumulated with the CLEO-c detector.
For $D^0$ and $D^+$ rates, we use 281 pb$^{-1}$ taken on the
$\psi(3770)$ resonance, and for $D_s^+$ rates, we use 195 pb$^{-1}$
taken at 4170 MeV. We find that the production rates of these
particles are larger in $D_s^+$ decays than in $D^0$ and $D^+$
decays. The $\phi$ rate, in particular, is 15 times greater. These
branching fractions can be used to measure $B_s$ yields either at
the $\Upsilon(5S)$ resonance or at hadron colliders.

\end{abstract}


\maketitle

\section{Introduction}
Inclusive decay rates of charm mesons into mesons with large
$s\overline{s}$ inherent quark content, the $\eta$, $\eta'$ and
$\phi$ mesons, are important for studies of both charm and $b$
decays. Non-strange $D$ mesons are generally expected not to decay
into such objects, while the $D_s^+$ is likely to have both $s$ and
$\overline{s}$ quarks present after the primary $c\to s W^+$
transition, resulting in many more such particles. This is
particularly useful in distinguishing between $B$ and $B_s$ mesons
as the $B$ decays into $D$'s with a large rate of $\sim$90\%, while
the decays into $D_s^+$ are at the $\sim$10\% level. We expect that
the reverse is true for $B_s$ mesons. Knowledge of the charm yields
into these mesons would allow alternative analyses of $B_s$ rates at
the $\Upsilon(5S)$ or at hadron colliders \cite{5S}.

In this analysis we use 281 pb$^{-1}$ integrated luminosity of
CLEO-c data produced in $e^+e^-$ collisions and recorded at the peak
of the $\psi''$ resonance (3.770 GeV) to study the $\eta$, $\eta'$
and $\phi$ yields in $D^0$ and $D^+$ decays. Production in $D_s^+$
decays is studied at 4170 MeV, where the cross-section for
$D_s^{*\pm}D_s^{\mp}$ is $\sim$1 nb \cite{poling}.

The CLEO-c detector is equipped to measure the momenta and
directions of charged particles, identify charged hadrons, detect
photons, and determine with good precision their directions and
energies. It has been described in more detail previously
\cite{CLEOD,CLEODR,RICH}.

\section{Selection of $D^0$, $D^-$ and $D_s^-$ Tagging Modes}
Fully reconstructed charged or neutral $D$ meson candidates are
selected from the data at 3.770 GeV, where pairs of
$D^0\overline{D}^0$ or $D^+D^-$ mesons are produced. The decay modes
used are listed in Table~\ref{tab:D0Dprecon}. In general, in this
paper, $D$ refers to either a $D^0$ or $D^+$ meson and its
anti-particle, and $D_s$ refers to $D_s^+$ meson and its
anti-particle. (Also, mention of a flavor specific state also
implies use of the charge-conjugate state.)

\begin{table}[htb]
\begin{center}
\begin{tabular}{llccc}
 &~     Mode                   &~ Signal          &~  Background \\
\hline
       &~ $K^+\pi^-\pi^- $            &~ $  77387 \pm 281 $    &~ $1868$\\
       &~ $K^+\pi^-\pi^- \pi^0$       &~ $  24850 \pm 214 $    &~ $12825$\\
 &~ $K_s\pi^-$                  &~ $  11162 \pm 136 $    &~ $514$\\
       &~ $K_s\pi^-\pi^-\pi^+ $       &~ $  18176 \pm 255 $    &~ $8976$\\
       &~ $K_s\pi^-\pi^0 $            &~ $  20244 \pm 170 $    &~ $5223$\\
\hline
       &~ Sum  $D^-$                        &~ $ 151819 \pm 487 $    &~ $29406$\\
\hline\hline
                    &~ $K^+\pi^- $             &~ $  49418 \pm 246 $    &~ $  630$\\
   &~ $K^+\pi^-\pi^0$         &~ $ 101960 \pm 476 $    &~ $18307$\\
                    &~ $K^+\pi^-\pi^+\pi^-$    &~ $  76178 \pm 306 $    &~ $ 6421$\\
\hline
                    &~ Sum $\overline{D}^0$                     &~ $ 227556 \pm 617 $    &~ $25357$\\
\hline\hline
\end{tabular}
\end{center}
\caption{Tagging modes and numbers of signal and background events
determined from the fits to the $D^-$ and $\overline{D}^0$ beam
constrained mass distributions, after making the mode-dependant
$m_{\mathrm{BC}}$ cuts. The error on the summed signal yield is
obtained by adding the errors on the individual yields in
quadrature.} \label{tab:D0Dprecon}
\end{table}

At 4.170 GeV  \cite{poling} we produce $D_s^+D_s^-$ pairs, with one
of the $D_s$ being, most of the time, the daughter of a $D_s^*$
decay. Fully reconstructed $D_s^+$ candidates are selected in the
decay modes listed in Table~\ref{tab:DSrecon}. $D$ mesons at this
energy are a source of background, they are mostly produced in
$D^{*}\overline{D}^*$ final states, with a cross-section of $\sim$5
nb, and $D^*\overline{D}+D\overline{D}^*$ final states, with a
cross-section of $\sim$2 nb. $D\overline{D}$ is a relatively small,
$\sim$0.2 nb. There also appears to be $D\overline{D}$ production
with extra pions.

\begin{table}[htb]
\begin{center}
\begin{tabular}{lccr}
~     ~~Mode                &~ Signal            &~  Background \\
\hline
        $K^+ K^- \pi^-$                &~ $ 8446 \pm 160 $   &~ 6793 \\
        $K_s K^- ~(K_s \to \pi^+\pi^-)$   &~ $ 1852 \pm 62 $    &~ 1022 \\
 $\eta\pi^-$                  &~ $ 1101 \pm 80 $    &~ 2803 \\
       $\eta'\pi^- $                  &~ $ 786 \pm 37 $     &~ 242 \\
        $\phi\rho^-$                  &~ $ 1140 \pm 59 $    &~ 1515 \\
        $K^{*+}(890){K^{*0}}(890)$ &~ $ 1197 \pm 81 $    &~ 2599 \\
\hline
       Sum                            &~ $ 14522 \pm 218 $  &~ 15328 \\

\hline\hline
\end{tabular}
\end{center}
\caption{Tagging modes and number of signal and background events
determined from the fits shown in Fig.~\ref{DStags}, after making
the mode dependant invariant mass cuts. The error on the summed
signal yield is obtained by adding the errors on the individual
yields in quadrature.} \label{tab:DSrecon}
\end{table}

We fully reconstruct one of the $D$ mesons at 3.770 GeV or one of
the $D_s$ mesons at 4.170 GeV to form a specific tag, and then look
for cases where the particle produced in association with our tag
has a decay of either $\eta\to\gamma\gamma$,
$\eta'\to\pi^+\pi^-\eta$, $\eta\to\gamma\gamma$, or $\phi\to
K^+K^-$.

All acceptable track candidates must have a helical trajectory
that approaches the event origin within a distance of 5 mm in the
azimuthal projection and 5 cm in the polar view, where the
azimuthal projection is in the bend view of the solenoidal magnet.
Each track must possess at least 50\% of the hits expected to be
on a track, and it must be within the fiducial volume of the drift
chambers, $|\cos\theta|<0.93$, where $\theta$ is the polar angle
with respect to the beam direction.

We reconstruct $\pi^0$'s by first selecting photon candidates from
energy deposits in the crystals that are not matched to charged
tracks and that have deposition patterns consistent with that
expected for electromagnetic showers. Pairs of photon candidates are
kinematically fit to the known $\pi^0$ mass \cite{PDG}.  We require
the pull, the difference between the reconstructed and known $\pi^0$
mass normalized by its uncertainty, to be less than three for
acceptable $\pi^0$ candidates.

 $K_S$ candidates are formed from a pair of charged pions that are constrained to
 come from a single vertex. We also require that the invariant mass of the two
 pions be within 4.5 times the width of the $K_S$ mass peak, which has an r.m.s. width of 4 MeV.

We use both charged particle ionization loss in the drift chamber
($dE/dx$) and Ring-Imaging Cherenkov (RICH) information to identify
kaons and pions used to fully reconstruct $D$ and $D_s$ mesons. The
RICH is used for momenta above 0.7 GeV/c. The angle of detected
Cherenkov photons that were radiated by a particular charged track
are translated into an overall likelihood denoted by ${\cal L}_i$
for each particle hypothesis. To differentiate between pion and kaon
candidates, we require the difference $-2\log({\cal
L_{\pi}})+2\log({\cal L}_K$) to be less than zero.  To utilize the
$dE/dx$ information, we calculate $\sigma_{\pi}$ as the difference
between the expected ionization loss for a pion and the measured
loss divided by the measurement error.  Similarly, $\sigma_{K}$ is
defined using the expected ionization for a kaon.

We use both the RICH and $dE/dx$ information for $D$ and $D_s$ meson
tag candidate tracks in the following manner: (a) If neither the
RICH nor $dE/dx$ information is available, then the track is
accepted as both a pion and a kaon candidate. (b) If $dE/dx$ is
available and RICH is not then we insist that pion candidates have
$PID_{dE}\equiv\sigma_{\pi}^2-\sigma_{K}^2 <0$, and kaon candidates
have $PID_{dE}> 0.$ (c) If RICH information is available and $dE/dx$
is not available, then we require that $PID_{RICH}\equiv
-2\log({\cal L}_{\pi})+2\log({\cal L}_K)<0$ for pions and
$PID_{RICH}>0$ for kaons. (d) If both $dE/dx$ and RICH information
are available, we require that $(PID_{dE}+PID_{RICH}) <0$ for pions
and $(PID_{dE}+PID_{RICH})>0$ for kaons.

\subsection{Reconstruction of $\overline{D}^0$ and $D^-$ Tagging Modes}
Tagging modes for $D^0$ and $D^+$ decays are reconstructed as
described previously \cite{CLEODptomunu}. Briefly, at the
$\psi(3770)$ $D$ meson final states are reconstructed by first
evaluating the difference, $\Delta E$, between the energy of the
decay products and the beam energy. We fit the $\Delta E$ spectrum
with a double Gaussian to represent the signal, and a polynomial
representing the background. We require the absolute value of this
difference to contain 98.8\% of the signal events, {\it i.~e.} to be
within $\sim$2.5 times the r.m.s width of the peak value. For final
states consisting entirely of tracks, the $\Delta E$ resolution is
$\sim 7~\rm MeV$. A $\pi^0$ in the final state degrades this
resolution by roughly a factor of two. Candidates with $\Delta E$
consistent with zero are selected, and then the $D$ beam-constrained
mass is evaluated,
\begin{equation}
m_{\mathrm{BC}}=\sqrt{E_{\mathrm{beam}}^2-(\sum_i\overrightarrow{p}_{\!i})^2},\label{eq:mBC}
\end{equation}
where $i$ runs over all the final state particles.

\begin{figure}[htb]
\centerline{ \epsfxsize=6in \epsfysize=6in
\epsffile{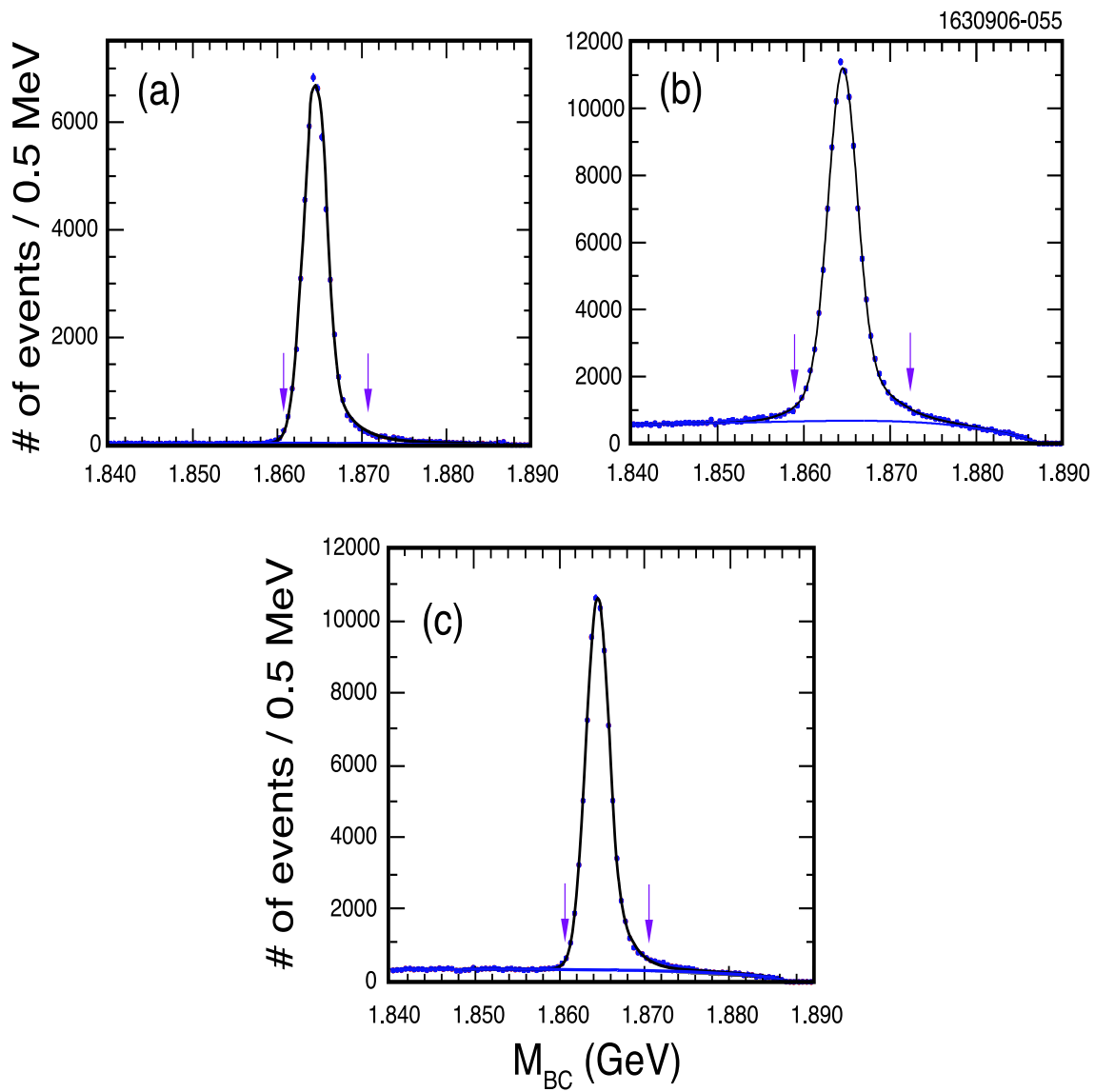} } \caption{\label{D0tags} Beam-constrained
mass distributions for fully reconstructed $\overline{D}^0$ decay
candidates in the final states: (a) $K^+ \pi^-$, (b) $K^+ \pi^-
\pi^0$, and (c) $K^+ \pi^-\pi^-\pi^+$. The distributions are fit to
a Crystal Ball Line shape for the signal. For the background, we
either use a fourth order polynomial (in (a) and (b)) or an ARGUS
shape (in (c)). Both background shapes are obtained from the $\Delta
E$ sidebands. The regions between the arrows are selected for
further analysis.}
\end{figure}

\begin{figure}[htbp]
\centerline{ \epsfxsize=6in
\epsffile{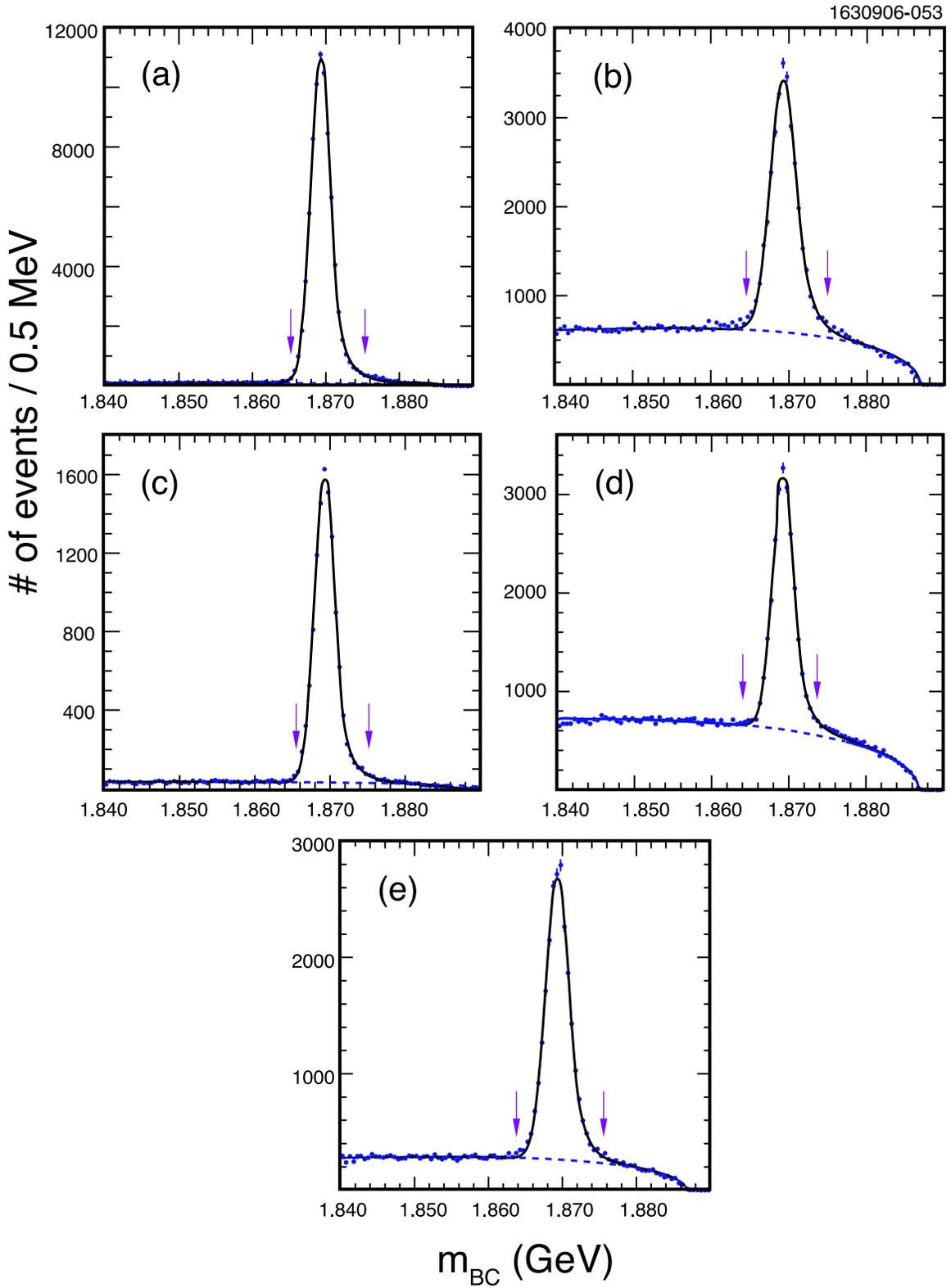} } \caption{\label{Dptags} Beam-constrained
mass distributions for fully reconstructed $D^-$ decay candidates in
the final states: (a) $K^+ \pi^- \pi^-$, (b) $K^+ \pi^- \pi^-
\pi^0$, (c) $K_s\pi^-$, (d) $K_s \pi^-\pi^+\pi^-$, and (e) $K_s\pi^-
\pi^0$. The distributions are fit to a Crystal Ball Line shape for
the signal and an ARGUS shape obtained from the $\Delta E$ sidebands
for the background. The regions between the arrows is selected for
further analysis.}
\end{figure}

The $m_{\mathrm{BC}}$ distributions for all $\overline{D}^0$ and
$D^-$ tagging modes are shown in Figs.~\ref{D0tags}
and~\ref{Dptags}. Table~\ref{tab:D0Dprecon} lists the numbers of
signal and background events within the signal region defined as
containing 98.8\% of the signal events with $m_{\mathrm{BC}}$ below
the peak and 95.5\% of the signal events above the peak; the
interval varies from mode to mode. The numbers of tagged events are
determined from fits of the $m_{\mathrm{BC}}$ distributions to a
signal function plus a background shape. The signal is described by
a Crystal Ball Line shape \cite {CBL,taunu}. For the background, we
fit with a shape function analogous to one first used by the ARGUS
collaboration \cite{ARGUS}, which has approximately the correct
threshold behavior at large $m_{\mathrm{BC}}$, except for the
$\overline{D}^0 \to K^-\pi^+$ and $\overline {D^0} \to
K^-\pi^+\pi^0$ modes where we use a fourth order polynomial. For
each tagging mode, the background function is first fit to a
$m_{\mathrm{BC}}$ distribution that lies within an interval from 5
to 7.5 r.m.s. widths away from the peak of the $\Delta E$
distribution. We fix the shape parameters from these fits and then
use these functions for background distributions in the signal fits,
allowing the normalization to float.

We find $151819\pm  487\pm 759$  $D^-$ and $ 227556\pm 617\pm 1138$
$\overline{D}^0$ signal events that we use for further analysis. The
systematic uncertainty on this number is estimated to be $\pm$0.5\%
by varying the fitting functions.

\subsection{Reconstruction of $D_s^-$ Tagging Modes}
\begin{figure}[htbp]
\includegraphics[width=74mm]{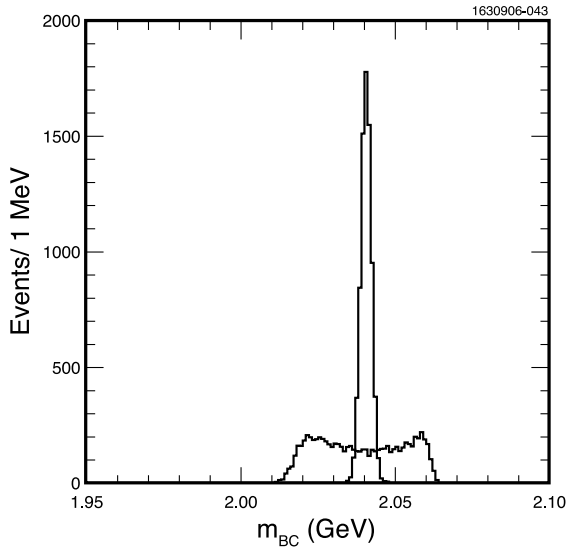}
\vspace{0.44mm}\caption{The beam constrained mass $m_{\mathrm{BC}}$
from Monte Carlo simulation of $e^+e^-\to D_s^+D_s^{*-}$,
$D_s^{*-}\to \gamma D_s^-$, $D_s^{\pm}\to\phi\pi^{\pm}$ at 4170 MeV.
The narrow peak is from the $D_s^+$ and the wider one from the
$D_s^-$. (The distributions are not centered at the $D_s^+$ or
$D_s^{*+}$ masses, because the reconstructed particles are assumed
to have the energy of the beam.)} \label{mbc}
\end{figure}

At 4170 MeV the presence of the $\gamma$ from $D_s^{*-}\to \gamma
D_s^{-}$ causes us to adopt a different procedure. If we ignore the
photon and reconstruct the $m_{\mathrm{BC}}$ distribution, we obtain
the distribution from Monte Carlo shown in Fig.~\ref{mbc}. The
narrow peak occurs when the reconstructed $D_s$ does not come from
the $D_s^*$ decay. Thus, the method of applying narrow cuts on
$m_{\mathrm{BC}}$ and $\Delta E$, used so successfully on the
$\psi(3770)$, no longer works.

\begin{figure}[htbp]
\centerline{ \epsfxsize=6.0in \epsffile{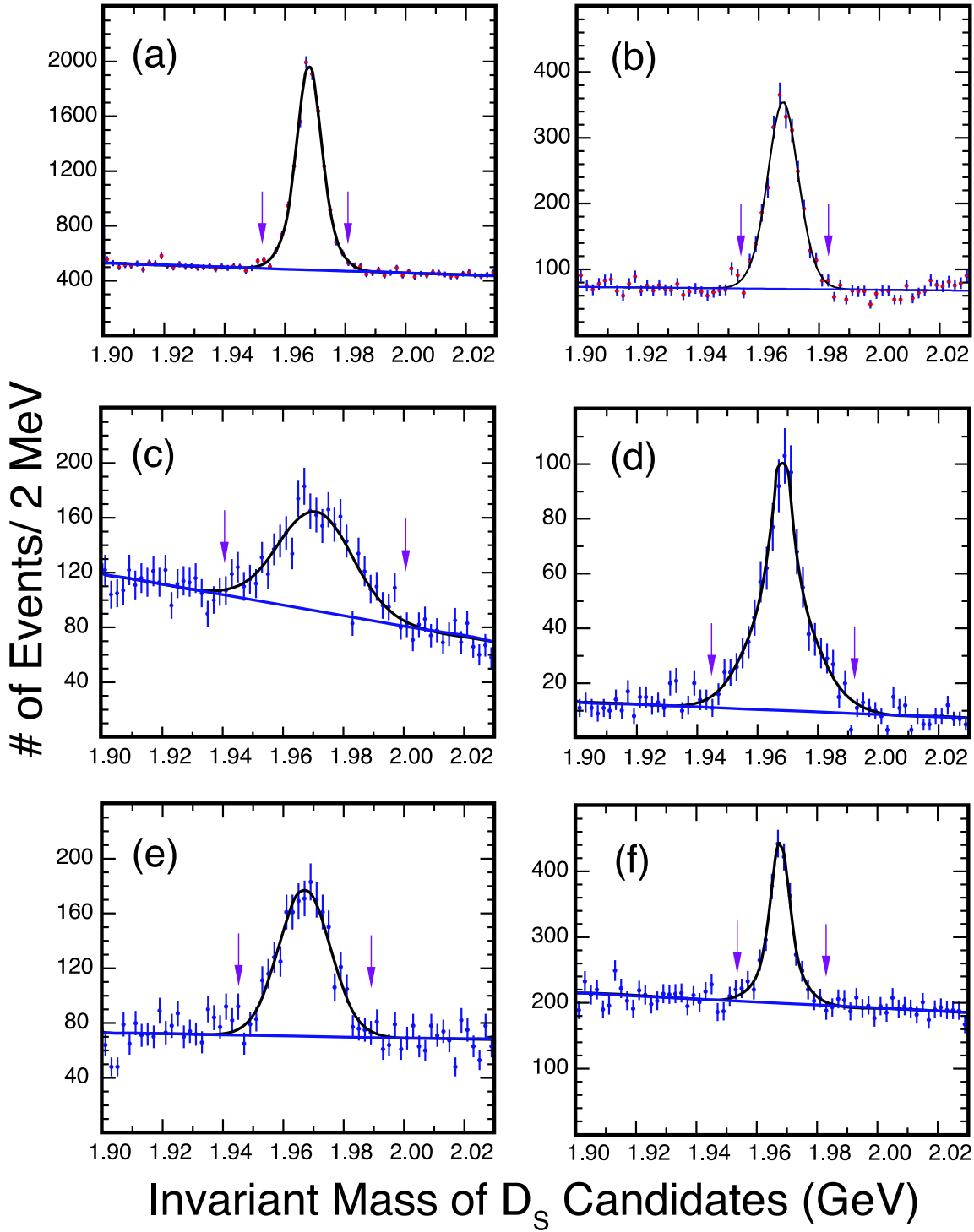} }
\caption{\label{DStags} Invariant mass distributions for fully
reconstructed $D_s^-$ decay candidates in the final states: (a) $K^+
K^- \pi^-$, (b) $K_s K^- $, (c) $\eta \pi^-$, (d) $\eta'\pi^-$, (e)
$\phi \rho^-$, and (f) $K^{*0}K^{*+}$. The distributions are fit to
double Gaussian signal shapes and Chebychev polynomial backgrounds.
The regions between the arrows is selected for further analysis.}
\end{figure}

Instead, we insist that the $D_s^{-}$ candidate has momenta which
satisfies the requirement 2.015 $< m_{\mathrm{BC}} <$ 2.067 GeV.
This requirement allows for the fact that the $D_s^{-}$ could have
been produced directly or as a result of a $D_s^{*-}$ decay to
either $\gamma D_s^-$, or $\pi^0 D_s^-$ decay with a small $\sim
5.8$\% branching fraction \cite{PDG}.
We then reconstruct the invariant mass of the $D_s$ candidates. The
invariant mass distributions of the tagging modes we considered in
this analysis are shown in Fig.~\ref{DStags}; they are listed in
Table~\ref{tab:DSrecon}, where we also give the number of signal and
background events. Here we use only the secondary decays
${K^{*0}}(890)\to K^+\pi^-$, ${K^{*+}}(890)\to K_s\pi^+$,
$\rho^-\to\pi^-\pi^0$, $\eta\to \gamma\gamma$ and
$\eta'\to\pi^+\pi^-\eta$. (More specifically, when appropriate, we
require the $K\pi$ invariant mass to be within $\pm$100 MeV of the
$K^*$ mass, the $\pi^-\pi^0$ mass to be within  $\pm$100 MeV of the
$\rho^-$ mass, the $\gamma\gamma$ invariant mass minus the $\eta$
mass divided by its error to be less than 3, and the invariant mass
of the $\gamma\gamma\pi^+\pi^-$ minus the $\gamma\gamma$ mass, for
$\gamma\gamma$ combinations consistent with the $\eta$ hypothesis,
to be within $\pm$10 MeV of the known $\eta'-\eta$ mass difference.)

 The
$D_s^-$ signal regions are defined as containing 98.8\% of the
signal events within an invariant mass window symmetric about the
$D_s^-$ mass peak. The intervals vary from mode to mode.  To find
the numbers of signal tag events, the invariant mass distributions
are fit to a sum of two Gaussian signal functions with the means
constrained to be the same and the r.m.s. widths allowed to float.
The background function is a second or third order Chebychev
polynomial.

We have $14522 \pm 218 \pm 145 $   $D_s^-$ signal events that we use
for further analysis. The systematic uncertainty is estimated to be
$\pm$1.0\% by varying the signal and background fitting functions.

\section{$\eta$, $\eta'$ and $\phi$ Selection}
\label{sec:sel}

For the $\eta$ we use the $\gamma\gamma$ final state, which has a
measured branching fraction of $(39.43 \pm 0.26)$\%. To detect
$\eta'$ we use the $\pi^+\pi^-\eta$ final state, which has a
branching fraction of $(44.3 \pm 1.5)$\%, with the $\eta$
subsequently decaying into $\gamma\gamma$. For the $\phi$ we use the
$K^+K^-$ final state with a rate of $(49.1\pm 0.6)$\% \cite{PDG}.

The track selection and particle identification requirements for the
signal side are identical to those for the tag side, except for
momenta less than $0.2~{\rm GeV/}c$, where we loosen the $dE/dx$
consistency requirement to $4~\sigma_{K}$. This is the case for both
$D$ and $D_s$ meson tags.

We accept photons only in the best-resolution region of the
detector, $|\cos\theta|<0.8$, where $\theta$ is the angle of the
photon with respect to the beam direction. Photon candidates must
not be matched to charged tracks, must have a reconstructed energy
greater than 30 MeV and have a spatial distribution in the crystals
consistent with that of an electromagnetic shower.

Candidates for $\eta'$ mesons are selected by combining $\eta$
candidates within 3 r.m.s. widths of the $\eta$ mass, with a $\pi^+$
and a $\pi^-$. The mass difference between $\eta\pi^+\pi^-$ and
$\eta$ is then examined. Both pions forming $\eta'$ and kaons
forming $\phi$ candidates are required to pass the track selection
and particle identification requirements. The signal yield is then
extracted from fits to the $\eta$, $\eta'-\eta$, and $\phi$ mass
distributions.

\section{Reconstruction Efficiencies}

\begin{figure}[htb]
\centerline{
\epsfxsize= 6.8in  
\epsffile{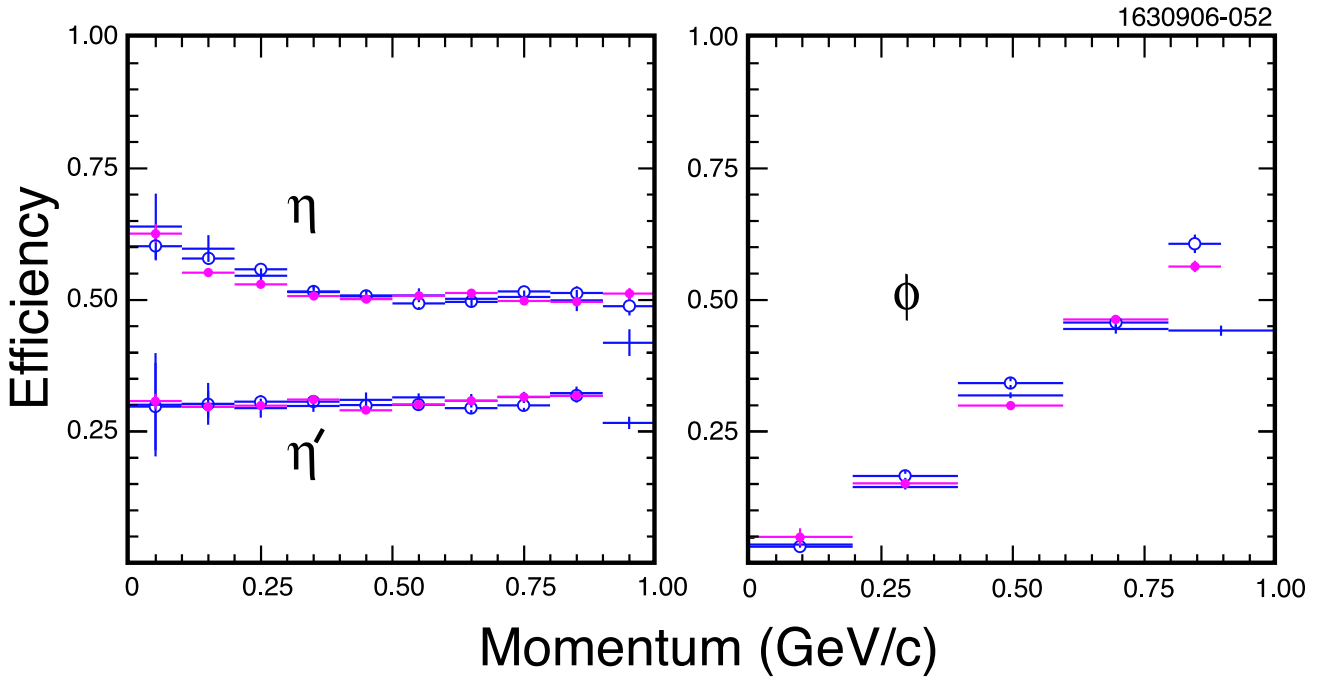}} \caption{ Reconstruction efficiencies for:
  $\eta \to \gamma\gamma$, $\eta' \to \eta\pi^+\pi^-$, and $\phi \to K^+K^-$.
The filled circles indicate $D^0\overline{D}^0$ events, the open
circles $D^+D^-$ events, and the crosses $D_s^*\overline{D}_s$
events. The efficiencies do not include branching ratios.}
\label{eff}
\end{figure}

The reconstruction efficiencies for $\eta$, $\eta'$ and $\phi$ in
our tag samples of $D$ and $D_s$ events are shown in Fig.~\ref{eff}.
They are determined from a Monte Carlo simulation of the detector
\cite{simulation}. There is no observable difference in the
efficiencies for $D^0$ and $D^+$ decays. In the case of the $D_s$,
the reconstruction efficiency, especially for the $\phi$, is lower
in the highest momentum bin, due to the different angular
distributions of these particles, caused by the different production
angles with respect to the beam of charmed mesons from
$D\overline{D}$ production compared with $D_s^*\overline{D}_s$
production. In other words, the kaons on average are produced closer
to the beam axis in $D_s^*\overline{D}_s$ events than in
$D\overline{D}$ events.

The $\eta$ efficiency falls slowly below 300 MeV/c, and then
levels out. Since our aim here is to measure the inclusive
branching fractions, we break the $\eta$ sample into two parts,
one below 300 MeV/c and the other above. For the $\eta'$, the
efficiency is constant with momentum, so we do not separate the
data into momentum intervals. The $\phi$ efficiency, on the other
hand, changes drastically with momentum and therefore we use
several momentum regions. The increase in the $\phi$ efficiency is
easily explained by the fact that as the $\phi$ becomes more
energetic it is less likely to produce a kaon of $p < 0.2
~\rm{GeV}/c$, which would cause the event to be rejected.

The simulated $\phi$ efficiency could be inaccurate if the $\phi$
polarization were not correct; this could occur because of a poor
choice of the mixture of final states. The data and Monte Carlo,
however, show the same polarization. (The observed polarization is
almost independent of momentum in both data and Monte Carlo.)

\section{Signal Yields and Branching Fractions}

The signal yields in this analysis are evaluated by taking the
difference between the $\eta$, $\eta'$ and $\phi$ yields opposite
selected tags and the yields in samples that estimate the background
under the tag peaks.
 Our procedure is somewhat different for $D$ and $D_s$
decays. In the $D$ case, we evaluate the background yields using
events in the low and high sidebands of the $\Delta E$ distribution
from $5\sigma$ to $7.5\sigma$ away from the peak. These sidebands
are normalized to have the same number of events as the backgrounds
under the $\Delta E$ peaks. In the $D_s$ case we select sidebands in
the same interval relative to the peak as for the $D$ but in
invariant mass rather than in $\Delta E$.

\subsection{Inclusive $\eta$ Yields}

\begin{figure}[htbp]
\centerline{ \epsfysize=6.5in \epsffile{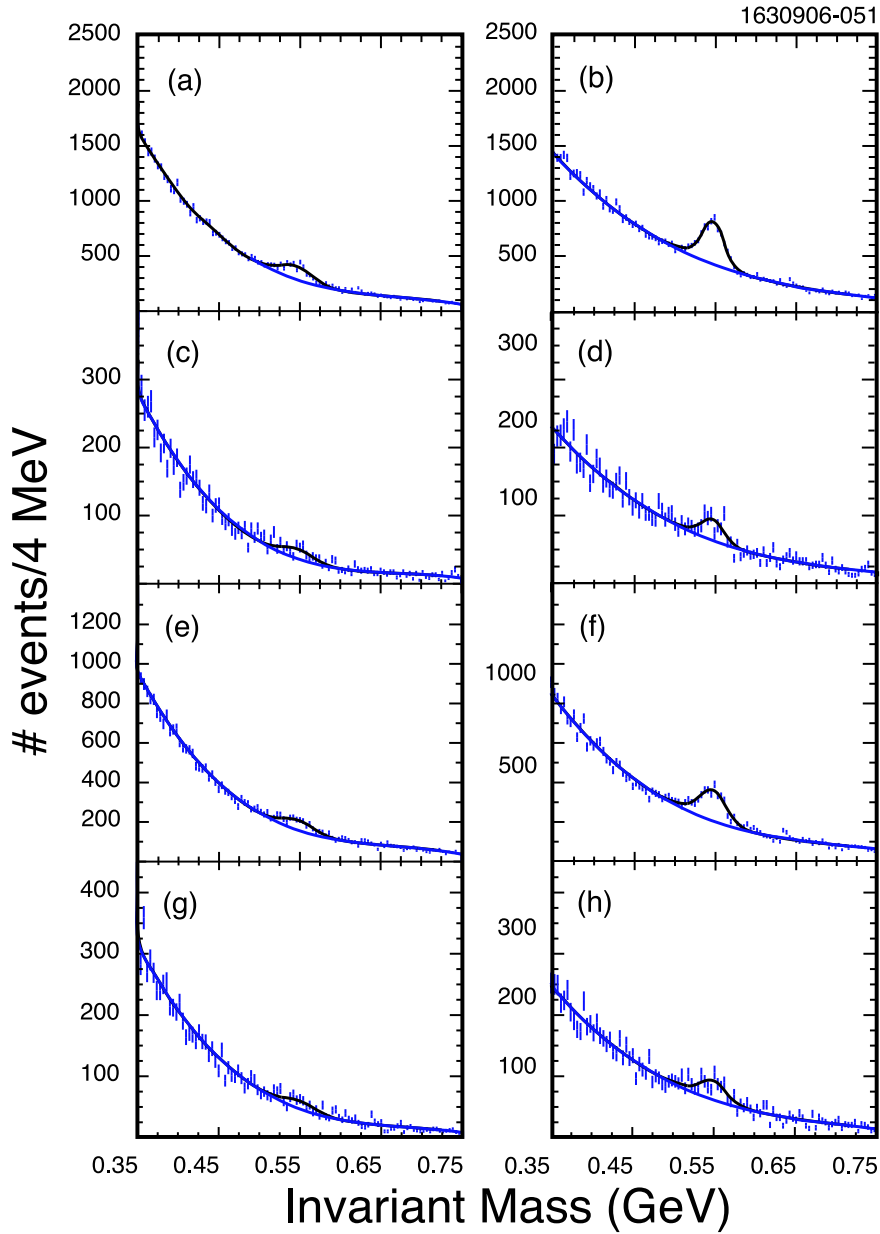} }
\caption{\label{eta_mass_d0} Invariant mass of the $\eta \to
\gamma\gamma$ candidates from $D^0$ decay: (a) signal region events
with the momentum of $\eta$, $|p_{\eta}|$, less than $0.3 ~{\rm
GeV/}c$, (b) signal region events with $0.3<|p_{\eta}|<1.0 ~{\rm
GeV/}c$, (c) sideband events with $|p_{\eta}|<0.3~{\rm GeV/}c$, (d)
sideband events with $0.3<|p_{\eta}|<1.0 ~{\rm GeV/}c$. Candidates
from $D^+$ decay are shown in (e)-(h), with corresponding
descriptions.}
\end{figure}

\begin{figure}[htb]
\centerline{ \epsfxsize=6.in \epsffile{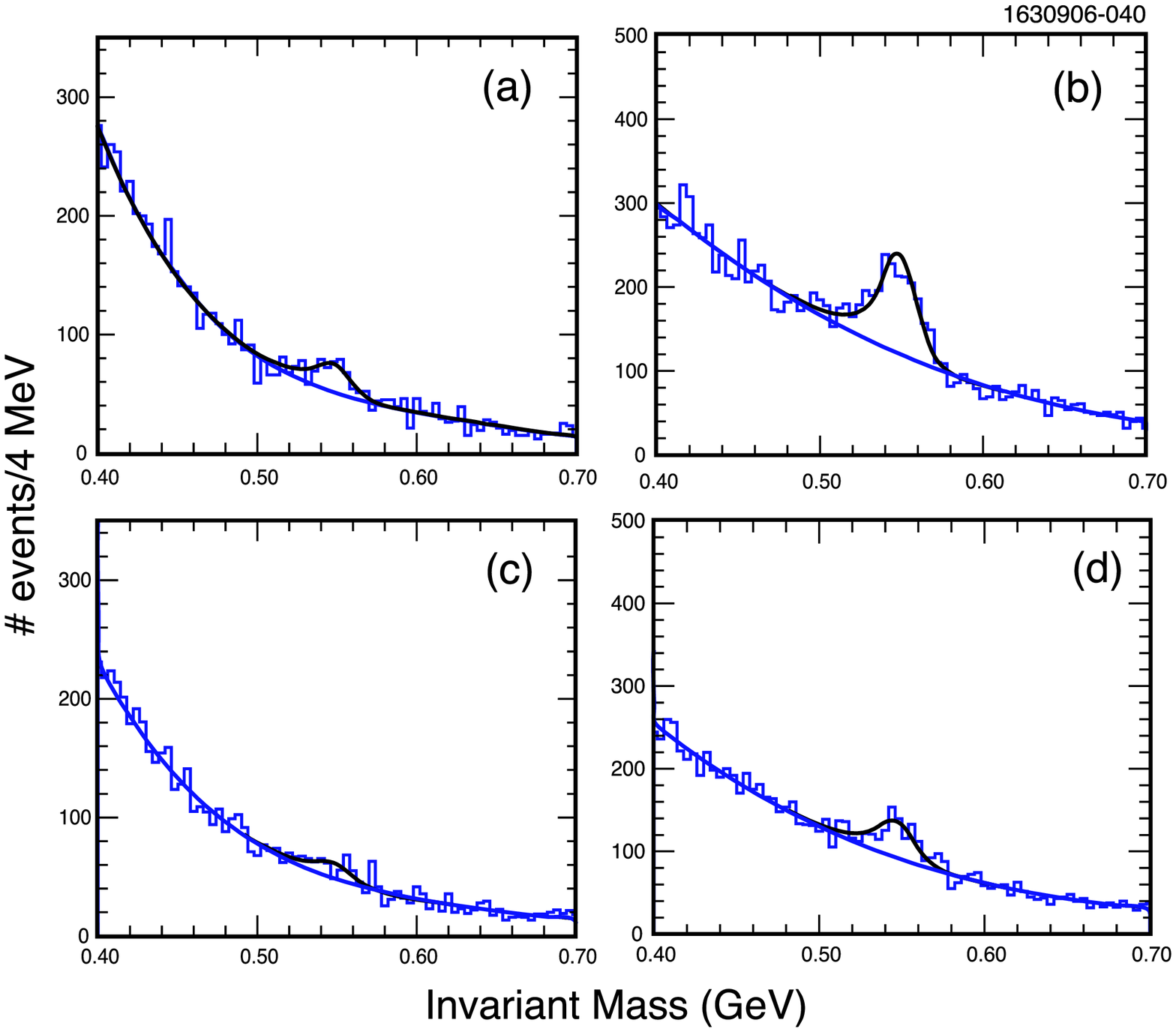} }
\caption{\label{ds-to-eta-sig-sb} Invariant mass of the $\eta \to
\gamma\gamma$ candidates from $D_s^+$ decay: (a) signal region
events with the momentum of $\eta$, $|p_{\eta}|$, less than $0.3
~{\rm GeV/}c$, (b) signal region events with $0.3<|p_{\eta}|<1.0
~{\rm GeV/}c$, (c) sideband events with $|p_{\eta}|<0.3~{\rm
GeV/}c$, (d) sideband events with $0.3<|p_{\eta}|<1.0 ~{\rm
GeV/}c$.}
\end{figure}

In Fig.~\ref{eta_mass_d0}  we show the
two-photon invariant mass in our two momentum intervals for both
signal and sideband regions for $D$ tags; Fig. \ref{ds-to-eta-sig-sb}
shows the corresponding distributions for $D_s$ tags.

The $\eta$ signal and background yields are determined by fits to a
Crystal Ball function, to account for the peak and the low mass
tail, and a background polynomial. For the signal $D^0$ region, the
four fit parameters describing the Crystal Ball line shape, three
for shape and one for the mean are allowed to float. These
parameters are then fixed and used for the $D^0$ sideband regions
and the $D^+$ and $D_s^+$ signal and sideband regions, since some of
these have relatively small yields. Table~\ref{tab:D0DptoEta} lists
the yields, the efficiencies, and the branching fractions for the
two momentum intervals. (Yields in the highest momentum bin include
a small number of events that slightly exceed 1 GeV/$c$.)

\begin{table}[htb]
\begin{center}
\begin{tabular}{lcccccc}
\hline Tag & $|p_{\eta}|~\rm$ & $N_{\eta}^{\mathrm{sig}}$ &
$N_{\eta}^{\mathrm{bkg}}$ & $N_{\eta}$ & $\epsilon^{i}$ &
${\cal{B}}^{i}(D\to \eta
X)$ \\
 &  (GeV/$c$) &  &  &  &  & $(\%)$
\\
\hline
$D^0$        & 0.0-0.3 & $1454\pm133$  & $176\pm 33$ & $1278\pm137$ & $57$ & $2.5\pm0.3$ \\
             & 0.3-1.0 & $3427\pm137$  & $242\pm 36$ & $3185\pm141$ & $50$ & $7.0\pm0.3$ \\
\hline
Sum& & $4880\pm191$  & $418\pm49$  & $4463\pm197$ & $ -  $ & $9.5\pm0.4$\\
 \hline
$D^+$        & 0.0-0.3 & $608\pm 65$   & $153\pm 35$ & $455\pm74$   & $58$ & $1.3\pm 0.2$ \\
             & 0.3-1.0 & $1811\pm 115$ & $294\pm39 $ & $1517\pm121$ & $51$ & $5.0\pm 0.4$ \\
\hline
Sum && $2419\pm 132$ & $448\pm332$ & $1972\pm142$ & $-   $ & $6.3\pm0.5$\\
\hline $D_s^+$& 0.0-0.3 & $230\pm 38$   & $152\pm 36$ & $78\pm53$   & $55$ & $2.5\pm 1.7$ \\
             & 0.3-1.0 & $963\pm 56$ & $367\pm48 $ &596$\pm74$ & $50$ & $21.0\pm 2.6$ \\
\hline
Sum& & $1193\pm 68$ & $519\pm 60$ &  $674\pm 91$ & $-  $& $23.5\pm 3.1$\\
 \hline\hline
\end{tabular}
\end{center}
\caption{$\eta$ signal yields ($N_{\eta}^{\mathrm{sig}}$),
background yields ($N_{\eta}^{\mathrm{bkg}}$) and
background-subtracted yields ($N_{\eta}$) as a function of momentum.
Also listed are the $\eta$ reconstruction efficiencies
($\epsilon^{i}$) in percent, and the partial branching fractions
versus momentum. (Yields in the highest momentum bin include a small
number of events that slightly exceed 1 GeV/$c$.)}
\label{tab:D0DptoEta}
\end{table}

The systematic uncertainties arise from several sources. For the
$\eta$ we estimate a detection efficiency error of $\pm$2\% per
photon\footnote{This is determined from a study of pions in
$\psi(2S) \to J/ \psi \pi^0 \pi^0$ transitions \cite{had_ichep}.}
for a total of $\pm$4\%. We also include an error due to fitting the
Monte Carlo samples. In addition, there is an uncertainty caused by
using the efficiency in only two momentum intervals, due to possible
variations in these intervals, amounting to 3\%. This error is
estimated by considering the effects of different parent momentum
distributions. The uncertainties on the tag yields are derived by
varying the fitting functions.  We also have significant
contributions from uncertainties on the signal and background
fitting function, determined by varying the functions. All the
systematic error contributions are listed in Table~\ref{tab:sys}.
For the $\eta'$ and $\phi$ modes we also list the estimated
uncertainties for finding the charged tracks and identifying their
species. These differ somewhat between the $\eta'$ and $\phi$
because of the different track momenta involved. For the $\phi$ mode
we include another additional error source due the lack of
efficiency in the first momentum bin that we will discuss in more
detail subsequently.

\begin{table}[htb]
\begin{center}
\begin{tabular}{lccc}
\hline Systematic uncertainties  &$\eta$(\%) & $\eta'$(\%)&$\phi$(\%) \\
\hline
Photon reconstruction &4.0 & 4.0 & -\\
Charged track finding & -& 1.4 & 5.0 \\
Particle identification & - & 2.0 & 2.0 \\
Monte Carlo fitting & 2.0 & 1.0 & 1.0 \\
Average efficiency & 3.0 & 3.2 &- \\
Number of tags ($D^0$ \& $D^+$)& 0.5 & 0.5 & 0.5\\
Number of tags ($D_s^+$) & 1.0 & 1.0 & 1.0\\
Signal \& Background Fitting & 6.5 & 5.3 & 2.1 \\
Estimate of 1st $p$ bin ($D^0$ \& $D^+$) & - & - & 2.0 \\
Estimate of 1st $p$ bin ($D_s^+$) & - & - & 3.1 \\
 \hline
Total ($D^0$ \& $D^+$)  &8.5 & 8.6 & 6.3\\
Total ($D_s^+$)  &8.5 & 8.6 & 6.8\\
\hline\hline
\end{tabular}
\end{center}
\caption{Systematic uncertainties ($\pm$) on the inclusive $\eta$,
$\eta'$ and $\phi$ branching ratios.} \label{tab:sys}
\end{table}

For the $\eta$ rates we find
\begin{eqnarray} {\cal B}(D^0\to\eta X) &=&(9.5\pm0.4\pm
0.8)\%\\\nonumber {\cal B}(D^+\to\eta X) &=& (6.3\pm0.5\pm
0.5)\%\\\nonumber {\cal B}(D_s^+\to\eta X) &=& (23.5\pm 3.1 \pm
2.0)\%.
\end{eqnarray}
Note that these rates naturally include cascade decays from
$\eta'\to\eta X$.

\subsection{Inclusive $\eta'$ Yields} 
\begin{figure}[htbp]
\centerline{ \epsfxsize=6.5in \epsfysize=6in \epsffile{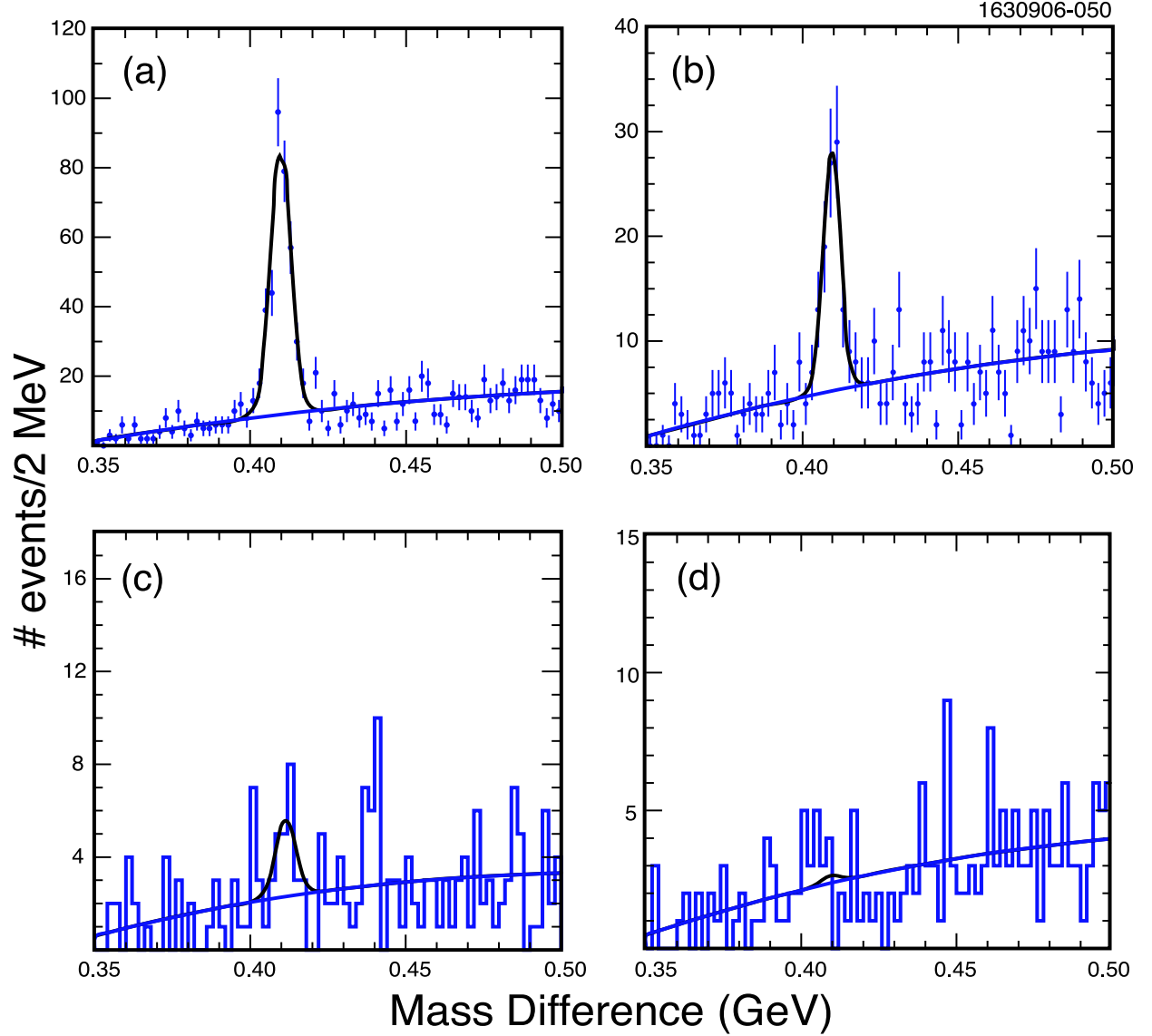}
} \caption{\label{diff_in_mass} Difference in the invariant mass of
$\eta' \to \eta\pi^+\pi^-$ and $\eta$ ($\eta \to \gamma\gamma$)
candidates from: (a) $D^0$ signal region events (b) $D^+$ signal
region events (c) $D^0$ sideband events, and (d) $D^+$ sideband
events. The fits are described in the text.}
\end{figure}

\begin{figure}[htbp]
\centerline{ \epsfxsize=6.5in \epsfysize=3.5in
\epsffile{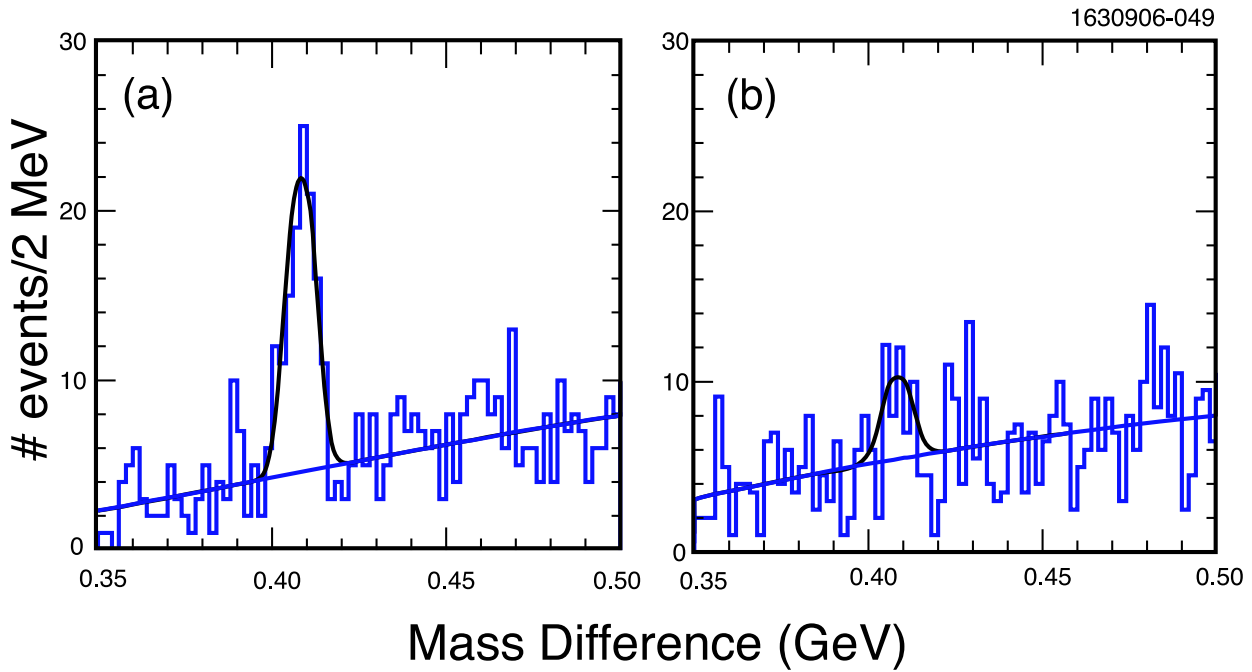} } \caption{\label{ds-to-etaprime-sig-sb}
$\eta\pi^+\pi^-$ - $\eta$ mass difference from $D^+_s{D^-_s}$ signal
events (a) and sideband events (b). The fits are described in the
text.}
\end{figure}

We first reconstruct
$\gamma\gamma$ invariant mass as shown above. Then we use
$\gamma\gamma$ mass combinations within $\pm3 \sigma$ of the $\eta$
mass as $\eta$ candidates, where $\sigma$ is the r.m.s. width of the
mass peak. In Figs.~\ref{diff_in_mass} and
\ref{ds-to-etaprime-sig-sb} we show the $\eta\pi^+\pi^-$ - $\eta$
mass difference for $D$ and $D_s$ tags from both signal and sideband
regions. To determine the yields in this case we fit to a Gaussian
signal function and a background polynomial. The signal shapes and
means are allowed to float for the signal distributions and fixed to
the values obtained there in the corresponding sideband regions.

The signal, background, and background-subtracted yields, the
detection efficiency and the branching fraction are given in
Table~\ref{tab:D0DptoEtaPrime}. The systematic error sources are
listed in Table~\ref{tab:sys}. For the charged tracks we estimate a
systematic error of $\pm$0.7\% for track finding\footnote{This is
determined from a study of pions in $\psi(2S) \to J/ \psi \pi^+
\pi^-$ transitions \cite{had_ichep} but increased to account for the
contribution from low momentum tracks.} and $\pm$1\% for particle
identification.
\begin{table}[htb]
\begin{center}
\begin{tabular}{lccccc}
\hline Tag &~ $N_{\eta'}^{\mathrm{sig}}$ &~
$N_{\eta'}^{\mathrm{bkg}}$ &~ $N_{\eta'}$ &~ $\epsilon^{i}$ &~
${\cal{B}}(D_{(s)}\to \eta'
X)(\%)$ \\
\hline
$D^0$ &~ $313\pm 20$ &~ $14\pm 5$ &~  $299\pm21$ &~ $ 30$ &~  $2.48\pm0.17\pm0.21$  \\
$D^+$ &~ $ 83\pm 12$ &~ $ 1^{+4}_{-1}$ &~  $ 82\pm13$ &~ $ 30$ &~  $1.04\pm0.16\pm0.09$  \\
$D_s^+$ &~ $ 91\pm 12$ &~ $ 23\pm 8$ &~  $ 68\pm 15$ &~ $ 31$ &~  $8.7\pm 1.9\pm 0.8$  \\
 \hline\hline
\end{tabular}
\end{center}

\caption{$\eta'$ signal yields ($N_{\eta'}^{\mathrm{sig}}$),
background yields ($N_{\eta'}^{\mathrm{bkg}}$) and
background-subtracted yields ($N_{\eta'}$), the $\eta'$
reconstruction efficiencies ($\epsilon^{i}$), and the branching
fractions.} \label{tab:D0DptoEtaPrime}
\end{table}

\subsection{Inclusive $\phi$ Yields}

\begin{table}[htb]
\begin{center}
\begin{tabular}{lccccc}
\hline
 $|p_{\phi}|~({\rm{ GeV}/}c)$ &~ $N_{\phi}^{\mathrm{sig}}$ &~
$N_{\phi}^{\mathrm{bkg}}$ &~
$N_{\phi}$ &~ $\epsilon^{i}$ (\%)&~ ${\cal{B}}^{i}(D_{(s)}\to \phi X)(\%)$ \\
\hline $D^0\to \phi X$ &&&&&\\
0.0 - 0.2 &~ $1.0\pm1.0$     &~ $1.0\pm1.0$   &~ $0.0\pm1.4$     &~ $5.2$  &~ $0.0\pm0.0$  \\
0.2 - 0.4 &~ $25.5\pm7.9$    &~ $2.0\pm1.4$   &~ $23.5\pm8.0$    &~ $15.3$ &~ $0.14\pm0.05$  \\
0.4 - 0.6 &~ $171.9\pm18.1$  &~ $3.2\pm2.8$   &~ $168.7\pm18.3$  &~ $30.2$ &~ $0.50\pm0.05$  \\
0.6 - 0.8 &~ $209.7\pm17.6$  &~ $11.3\pm3.8$  &~ $198.4\pm18.0$  &~ $46.5$ &~ $0.38\pm0.03$  \\
0.8 - 0.9 &~ $8.7\pm3.5$     &~ $1.0\pm1.0$   &~ $7.7\pm3.7$     &~ $56.6$ &~ $0.012\pm0.006$  \\
\hline
Sum &~ $416.7\pm26.7$  &~ $18.5\pm5.1$  &~ $398.2\pm27.2$  &~ $-$    &~ $1.03\pm0.08$  \\
\hline $D^+\to \phi X$ &&&&& \\
0.0 - 0.2 &~ $3.0\pm1.7$      &~ $1.0\pm1.0$  &~ $2.0\pm2.0$    &~ $3.3$  &~ $0.08\pm0.08$   \\
0.2 - 0.4 &~ $49.9\pm8.6$     &~ $7.7\pm3.3$  &~ $42.2\pm9.2$   &~ $16.8$ &~ $0.34\pm0.07$   \\
0.4 - 0.6 &~ $90.2\pm11.8$    &~ $12.1\pm4.2$ &~ $78.2\pm12.5$  &~ $34.4$ &~ $0.30\pm0.05$   \\
0.6 - 0.8 &~ $127.6\pm14.0$   &~ $7.6\pm3.0$  &~ $119.9\pm14.4$ &~ $45.9$ &~ $0.35\pm0.04$   \\
0.8 - 0.9 &~ $6.8\pm3.1$      &~ $1.0\pm1.0$  &~ $5.8\pm3.3$    &~ $60.9$ &~ $0.013\pm0.002$  \\
\hline
Sum &~ $277.4\pm20.6$   &~ $29.4\pm6.3$ &~ $248.0\pm21.3$ &~ $-$    &~ $1.08\pm0.12$  \\
\hline $D_s^+\to \phi X$&&&&& \\
0.0 - 0.2  &~  $  1.4\pm 1.6$   &~  $ 1.2\pm1.4$   &~  $  0.1\pm 2.1$  &~ $ 3.7$   &~ $0.1\pm0.8$  \\
0.2 - 0.4  &~  $ 49.0\pm 7.4$   &~  $ 8.3\pm3.5$   &~  $ 40.7\pm 8.2$  &~ $14.6$   &~ $3.9\pm0.8$  \\
0.4 - 0.6  &~  $144.4\pm12.9$   &~  $28.4\pm6.2$   &~  $116.1\pm14.3$  &~ $32.1$   &~ $5.1\pm0.6$  \\
0.6 - 0.8  &~  $155.6\pm13.1$   &~  $20.3\pm5.1$   &~  $135.3\pm14.1$  &~ $44.7$   &~ $4.2\pm0.5$  \\
 $>$0.8    &~  $ 82.1\pm 9.1$   &~  $ 6.3\pm3.0$   &~  $ 75.8\pm 9.6$  &~ $44.4$   &~ $2.4\pm0.3$  \\
\hline
  Total    &~  $ 432.5\pm21.9$  &~  $64.5\pm9.4$   &~ $368.0\pm23.8$   &~ $-$      &~ $15.7\pm1.4$  \\
\hline\hline
\end{tabular}
\end{center}
\caption{Measured $\phi$ signal yields ($N_{\phi}^{\mathrm{sig}}$),
background yields ($N_{\phi}^{\mathrm{bkg}}$) and
background-subtracted yields ($N_{\phi}$) versus momentum from $D$
and $D_s$ decays. Also listed are the $\phi$ reconstruction
efficiencies ($\epsilon^{i}$), and the partial branching fractions
vs momentum. (Note that measurements in the lowest momentum interval
will be replaced by a model dependent estimate.)} \label{tab:DtoPhi}
\end{table}

\begin{figure}[htb]
\centerline{\epsfxsize=5.5in \epsffile{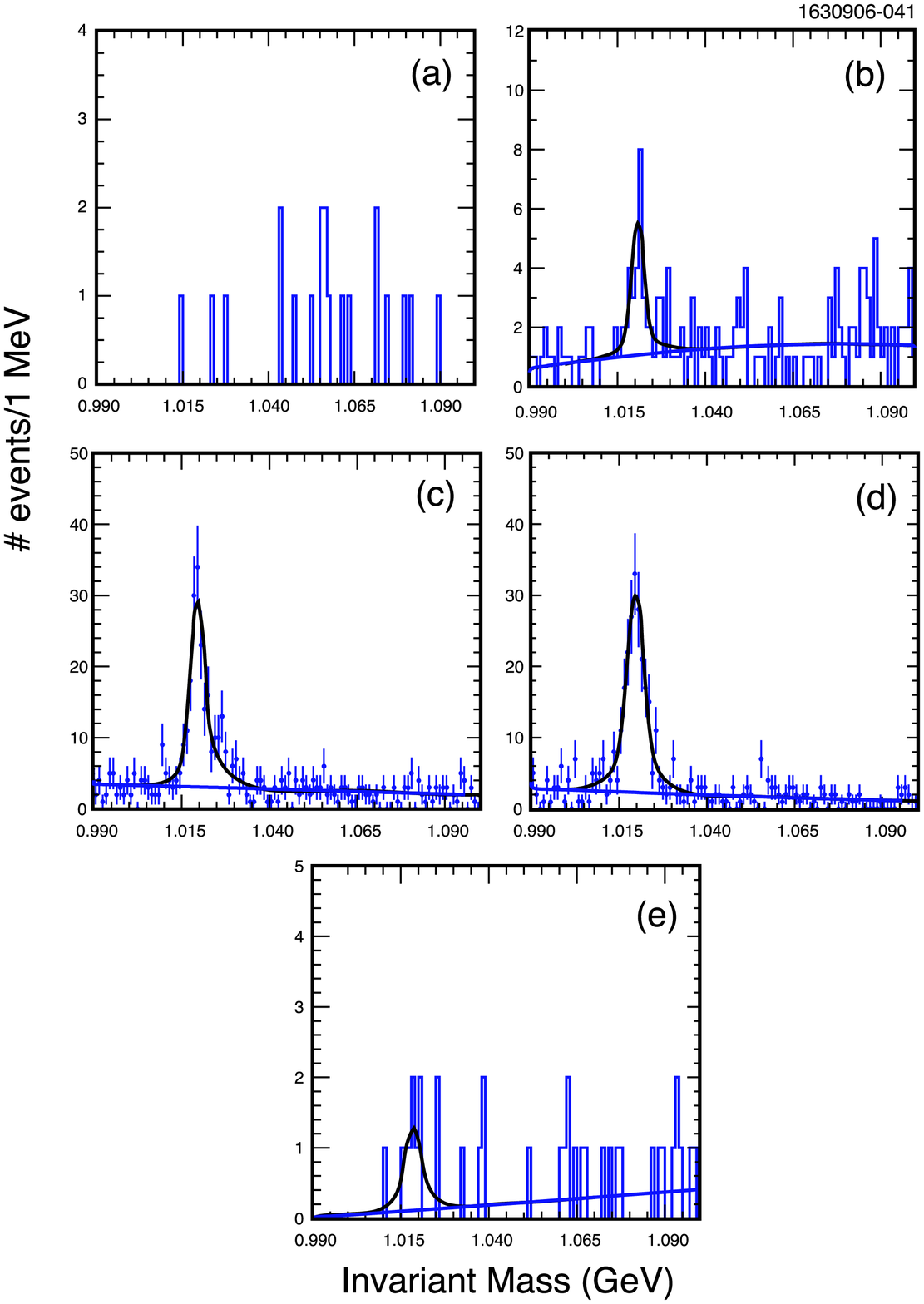} }
\caption{\label{phi-mass-d0-signal} Invariant mass of $\phi \to
K^+K^-$ candidates from $D^0\overline{D}^0$ signal events in five
different momentum intervals: (a) $0<|p_{\phi}|<0.2$~GeV/$c$, (b)
$0.2<|p_{\phi}|<0.4$~GeV/$c$, (c) $0.4<|p_{\phi}|<0.6$~GeV/$c$, (d)
$0.6<|p_{\phi}|<0.8$~GeV/$c$, (e) $0.8<|p_{\phi}|<0.9$~GeV/$c$. The
fits are described in the text.}
\end{figure}

\begin{figure}[htbp]
\centerline{\epsfxsize=5.5in \epsffile{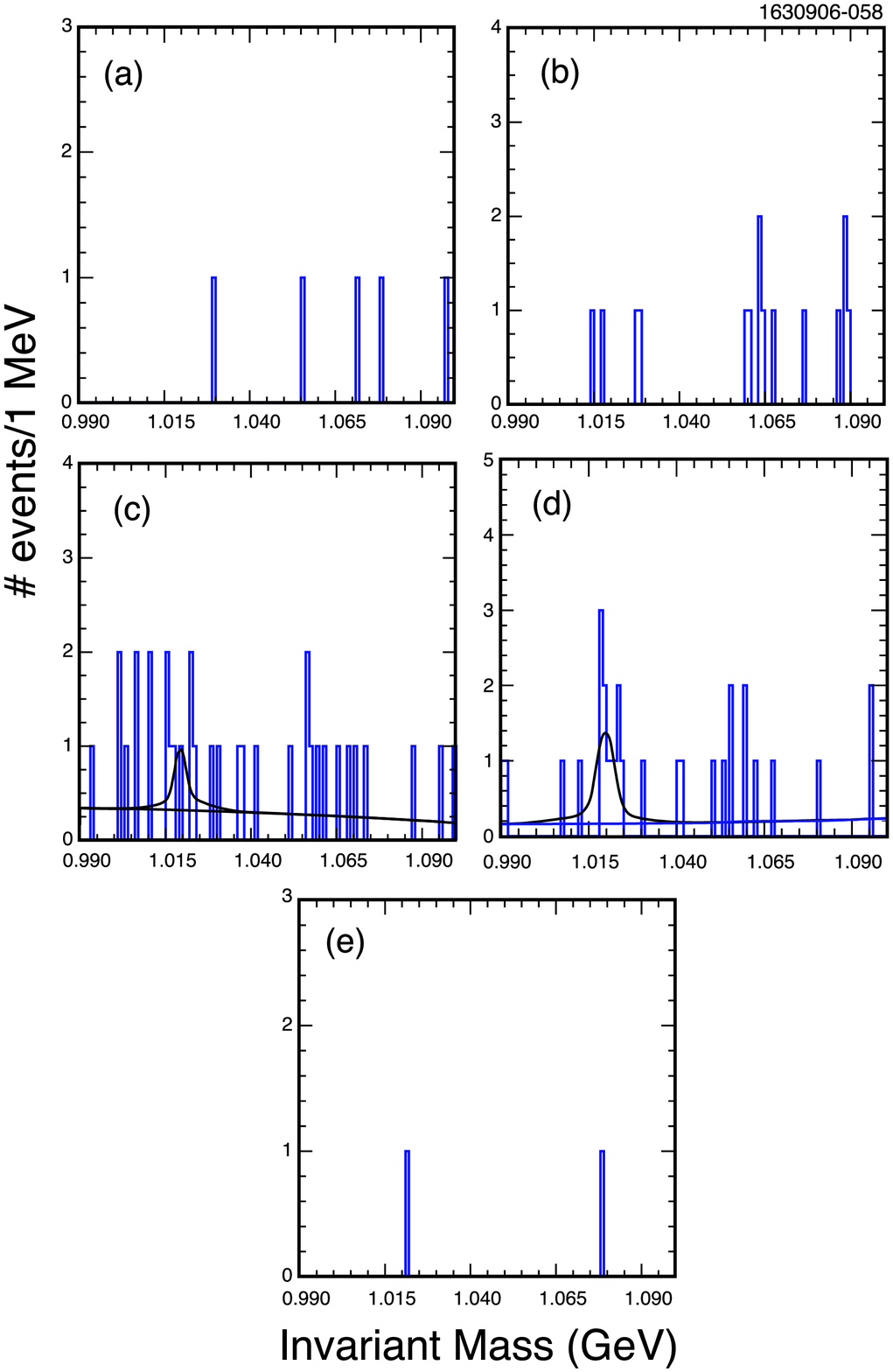} }
\caption{\label{phi-mass-d0-sb} Invariant mass of $\phi \to K^+K^-$
candidates from $D^0\overline{D}^0$ background events in five
different momentum intervals: (a) $0<|p_{\phi}|<0.2~{\rm{ GeV/}}c$,
(b) $0.2<|p_{\phi}|<0.4~{\rm{ GeV/}}c$, (c)
$0.4<|p_{\phi}|<0.6~{\rm{ GeV/}}c$, (d) $0.6<|p_{\phi}|<0.8~{\rm{
GeV/}}c$, (e) $0.8<|p_{\phi}|<0.9~{\rm{ GeV/}}c$. The fits are
described in the text.}
\end{figure}

\begin{figure}[htbp]
\centerline{ \epsfxsize=5.5in \epsffile{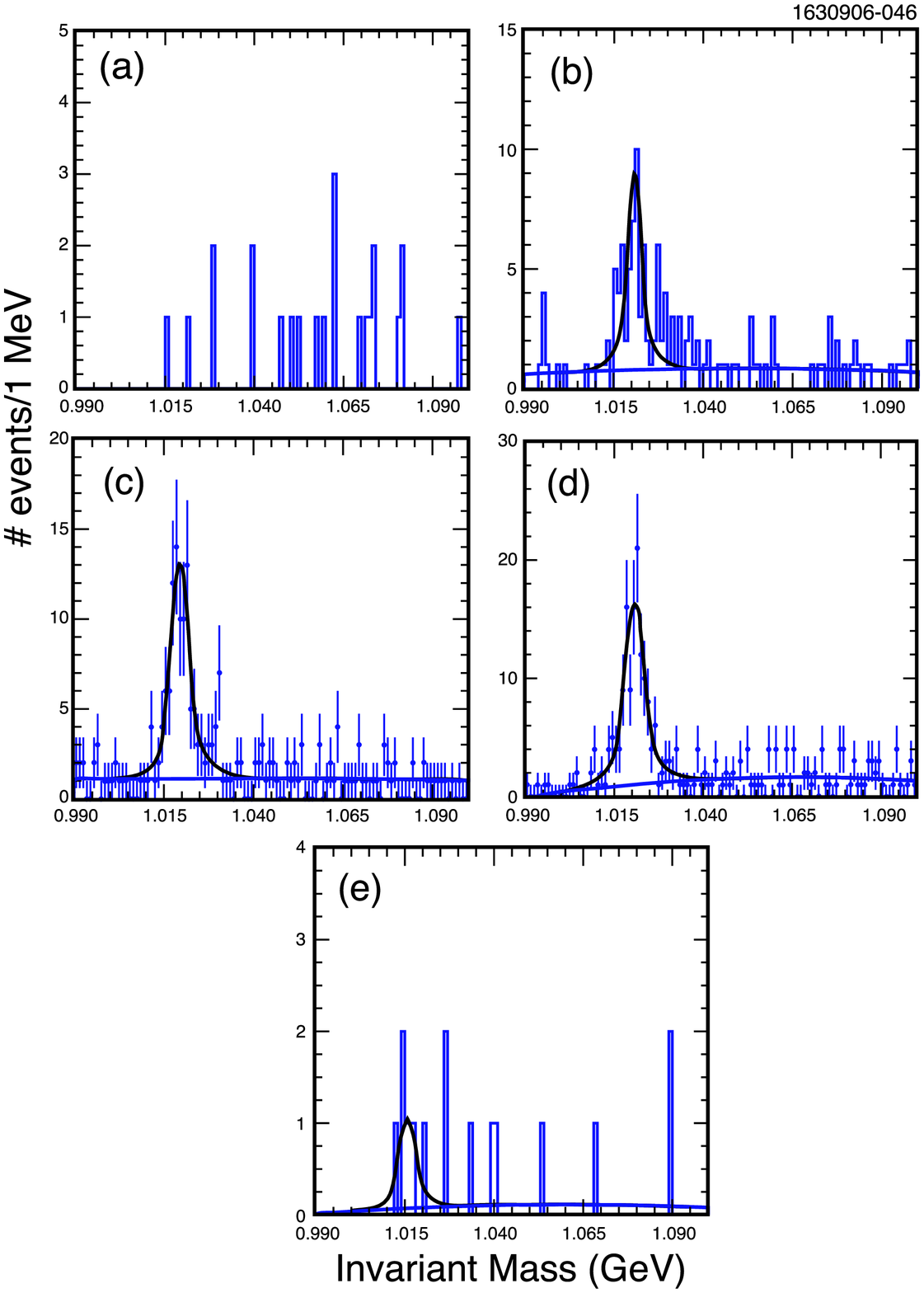} }
\caption{\label{phi-mass-dp-signal} Invariant mass of $\phi \to
K^+K^-$ candidates from $D^+{D^-}$ signal events in five different
momentum intervals: (a) $0<|p_{\phi}|<0.2~{\rm{ GeV}/}c$, (b)
$0.2<|p_{\phi}|<0.4~{\rm{ GeV}/}c$, (c) $0.4<|p_{\phi}|<0.6~{\rm{
GeV}/}c$, (d) $0.6<|p_{\phi}|<0.8~{\rm{ GeV}/}c$, (e)
$0.8<|p_{\phi}|<0.9~{\rm{ GeV}/}c$. The fits are described in the
text.}
\end{figure}

\begin{figure}[htbp]
\centerline{\epsfxsize=5.5in \epsffile{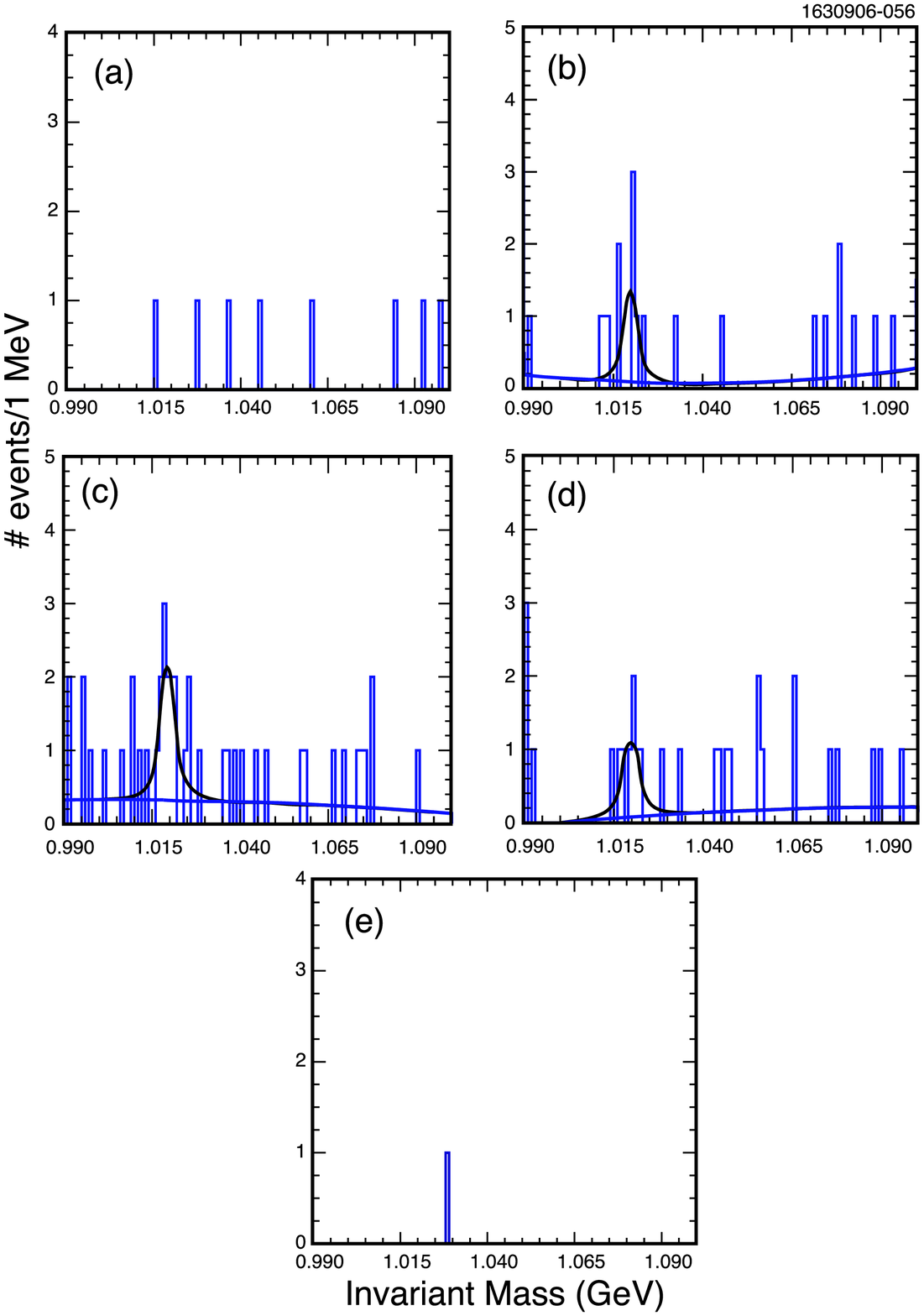} }
\caption{\label{phi-mass-dp-sb} Invariant mass of $\phi \to K^+K^-$
candidates from $D^+D^-$ background events in five different
momentum intervals: (a) $0<|p_{\phi}|<0.2~{\rm{ GeV/}}c$, (b)
$0.2<|p_{\phi}|<0.4~{\rm{ GeV/}}c$, (c) $0.4<|p_{\phi}|<0.6~{\rm{
GeV/}}c$, (d) $0.6<|p_{\phi}|<0.8~{\rm{ GeV/}}c$, (e)
$0.8<|p_{\phi}|<0.9~{\rm{ GeV/}}c$. The fits are described in the
text.}
\end{figure}

\begin{figure}[htbp]
\centerline{ \epsfxsize=5.5in \epsffile{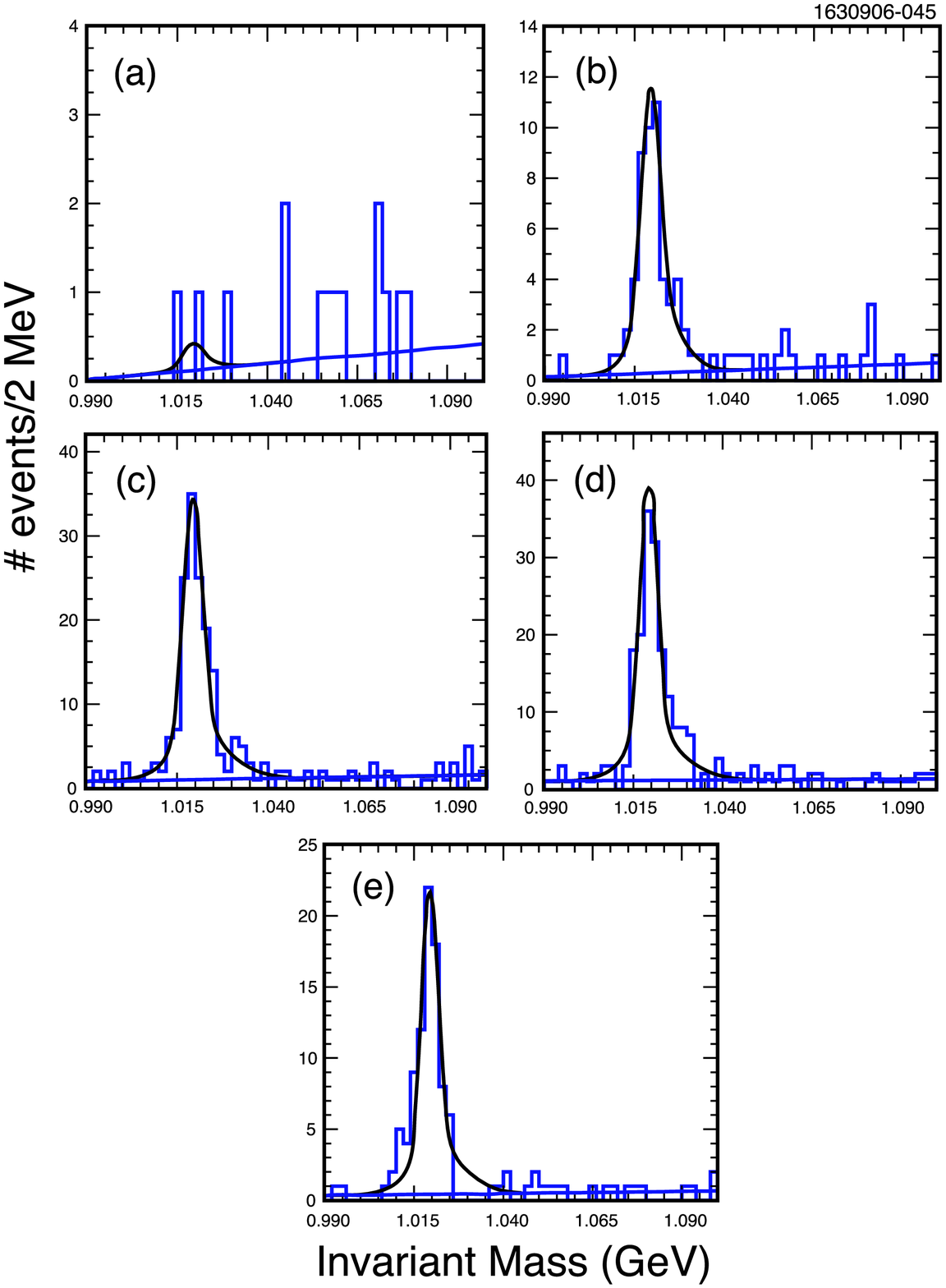} }
\caption{\label{ds-to-phi-sig} Invariant mass of $\phi \to K^+ K^-$
candidates from $D^+_s{D^-_s}$ signal events in five different
momentum intervals: (a) $0<|p_{\phi}|<0.2~{\rm{ GeV}/}c$, (b)
$0.2<|p_{\phi}|<0.4~{\rm{ GeV}/}c$, (c) $0.4<|p_{\phi}|<0.6~{\rm{
GeV}/}c$, (d) $0.6<|p_{\phi}|<0.8~{\rm{ GeV}/}c$, (e)
$|p_{\phi}|>0.8~{\rm{ GeV}/}c$. The fits are described in the text.}
\end{figure}

\begin{figure}[htbp]
\centerline{\epsfxsize=5.5in \epsffile{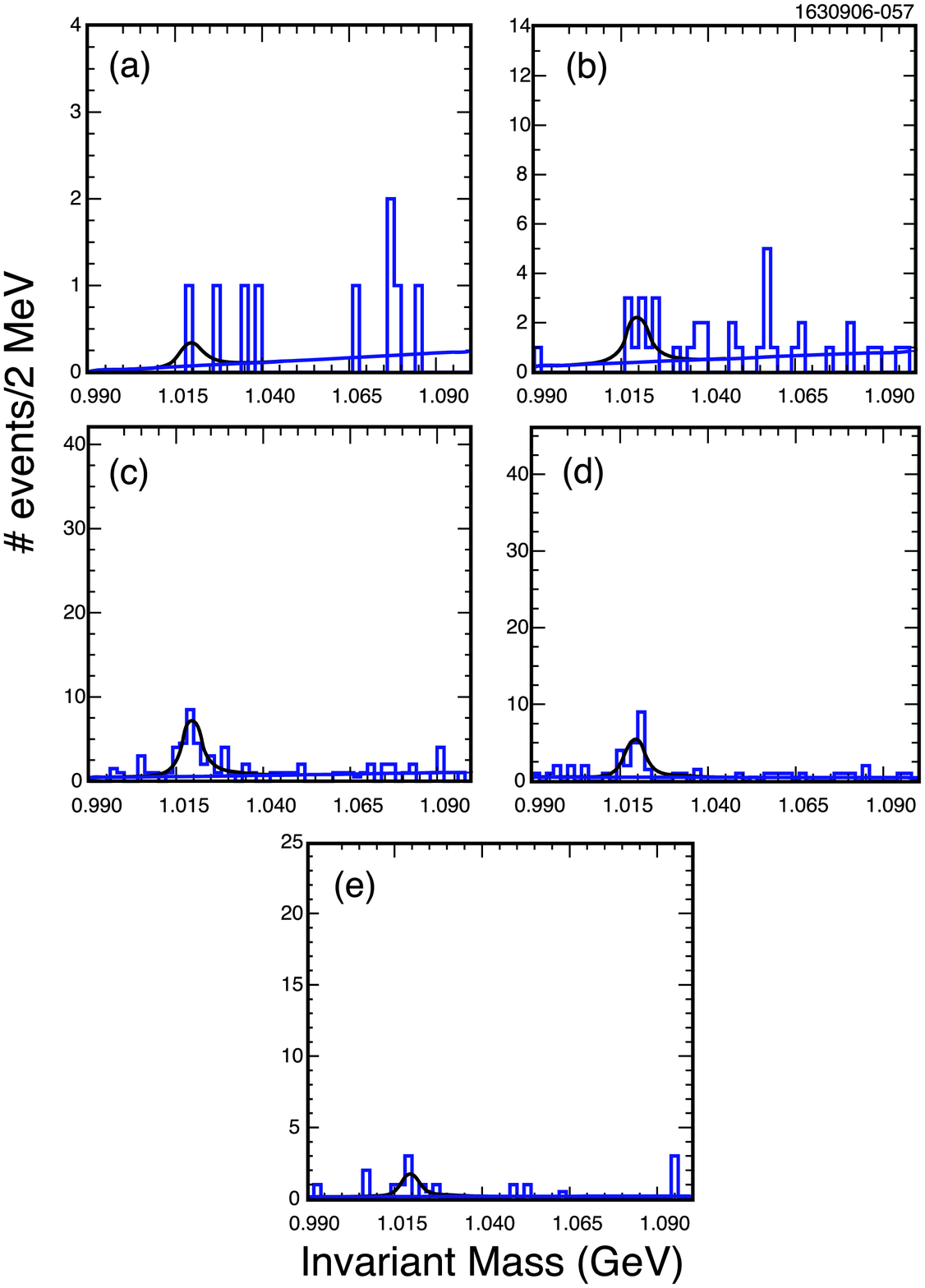} }
\caption{\label{ds-to-phi-sb} Invariant mass of $\phi \to K^+K^-$
candidates from $D^+_s{D^-_s}$ background events in five different
momentum intervals: (a) $0<|p_{\phi}|<0.2~{\rm{ GeV/}}c$, (b)
$0.2<|p_{\phi}|<0.4~{\rm{ GeV/}}c$, (c) $0.4<|p_{\phi}|<0.6~{\rm{
GeV/}}c$, (d) $0.6<|p_{\phi}|<0.8~{\rm{ GeV/}}c$, (e)
$0.8<|p_{\phi}|~{\rm{ GeV/}}c$. The fits are described in the text.}
\end{figure}

In Figs.~\ref{phi-mass-d0-signal}-\ref{ds-to-phi-sb} we show the
$K^+K^-$ invariant mass for the signal region in five different
momentum intervals from both signal and sideband regions for $D^0$,
$D^+$ and $D_s^+$ tags respectively. The signals are fit with a sum
of two Gaussian shapes and the background is fit to a polynomial.
The signal shapes are fixed to the values obtained by Monte Carlo
simulation, while the mean is allowed to float. The signal,
background, and background-subtracted yields, the detection
efficiency and the branching fraction in each momentum interval are
given in Table~\ref{tab:DtoPhi}. For $D^0$ and $D^+$ there are no
events above 0.9 GeV/$c$, while for the $D_s^+$ there are a small
number of events above 1.0 GeV/$c$.

\begin{figure}[htb]
\centerline{ \epsfxsize=6in \epsffile{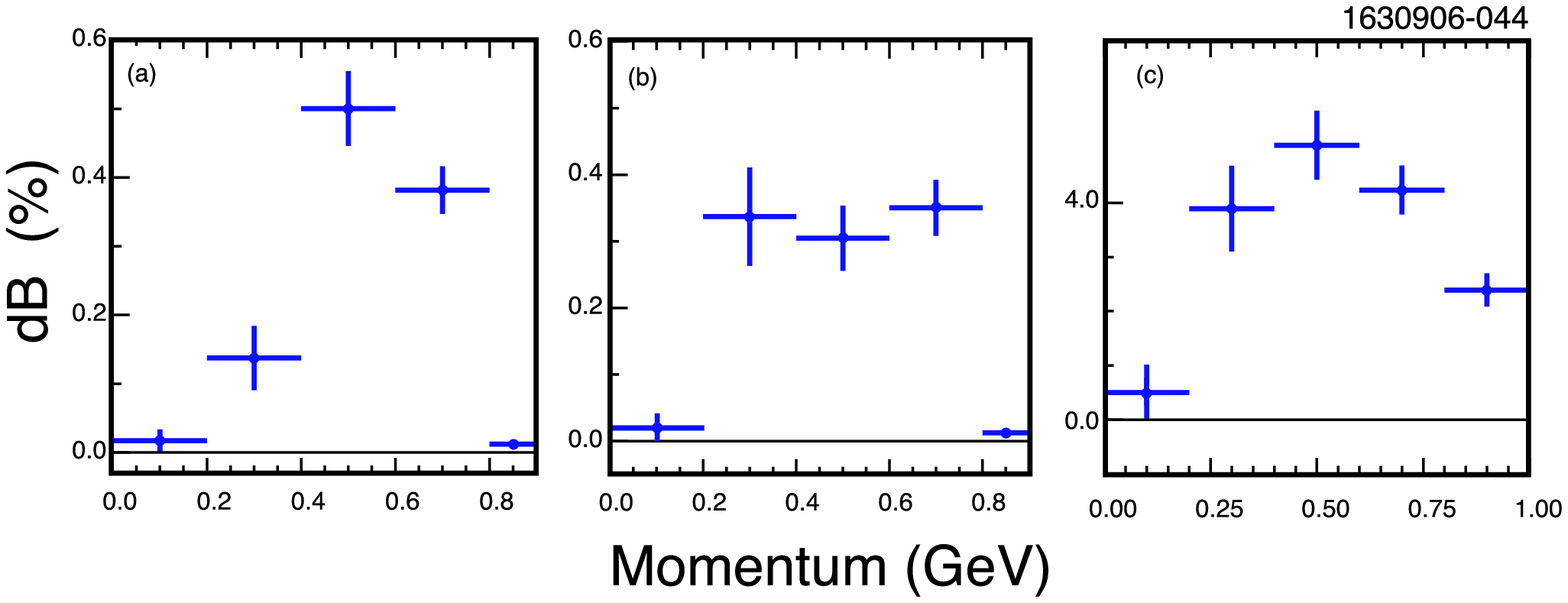} } \caption{
\label{PhiBr} The branching fraction in \% in each 200 MeV interval
of momentum for $\phi$ mesons from (a) $D^0$ decays, (b) $D^+$
decays and (c) $D_s^+$ decays.}
\end{figure}

Although the measured yields in the lowest momentum bin,
0$<p<$0.2~GeV/$c$, are quite small, so are the efficiencies. To take
into account possible incorrect efficiency estimates in this
difficult kinematic region we make an independent estimate of the
rate by using the Monte Carlo predicted fraction of the $\phi$
yield. We then take a conservative 100\% error on these estimates.
The fractions are 2.0\%, 2.0\% and 3.1\%, for $D^0$, $D^+$ and
$D_s^+$, respectively. These correspond to partial branching
fractions of (0.02$\pm$0.02)\%, (0.02$\pm$0.02)\%, and
(0.5$\pm$0.5)\%, respectively. Using these more reliable estimates
for the rates in the first bin, we show the efficiency corrected
momentum distributions in Fig.~\ref{PhiBr}. The inclusive branching
ratios are
\begin{eqnarray}
{\cal B}(D^0\to\phi X) &=& (1.05\pm0.08\pm0.07)\%\\\nonumber
 {\cal
B}(D^+\to\phi X) &=& (1.03\pm0.10\pm 0.07)\%\\\nonumber
 {\cal
B}(D_s^+\to\phi X) &=& (16.1\pm 1.2\pm 1.1)\%~.\nonumber
\end{eqnarray}
The systematic uncertainties are listed in Table~\ref{tab:sys}. Due
to the difficulty in tracking slow charged kaons, we use an
increased systemic error of $\pm$2.5\% per track in evaluating the
efficiency error from track finding. The error due to particle
identification remains at $\pm$1\%.

\section{Conclusions}

Our results of the inclusive $\eta$, $\eta'$ and $\phi$ production
rates from $D^0$, $D^+$ and $D_s^+$ decays are summarized in
Table~\ref{tab:sum}. Of the 9 measured rates in this paper, 8 are
first measurements and the other one, $D^0\to\phi X$, improves the
accuracy from 50\% to 10\%. We are consistent with previous upper
limits \cite{PDG} in the three cases where they exist.

\begin{table}[htb]
\begin{center}
\begin{tabular}{lcccccc}
\hline
Mode ~& \multicolumn{2}{c}{~~~~$D^0$(\%)} ~& \multicolumn{2}{c}{$D^+$(\%)} ~& \multicolumn{2}{c}{$D_s^+$(\%)} \\\hline
           &~ Our result            &~ PDG         &~ Our result           &~ PDG    &~ Our result     &~ PDG \\
\hline
$\eta$ $X$   &~  9.5$\pm$0.4$\pm$0.8   &~ $<$13       &~ 6.3$\pm$0.5$\pm$0.5 &~ $<$13  &~ 23.5$\pm3.1\pm2.0 $ &~ -  \\
$\eta'$ $X$  &~  2.48$\pm$0.17$\pm$0.21   &~ -           &~ 1.04$\pm$0.16$\pm$0.09 &~ -      &~ 8.7$\pm1.9\pm 0.8 $  &~ -  \\
$\phi$ $X$   &~  1.05$\pm$0.08$\pm$0.07   &~ 1.7$\pm$0.8 &~ 1.03$\pm$0.10$\pm$0.07 &~ $<$1.8 &~ 16.1$\pm 1.2\pm 1.1 $  &~ -  \\
 \hline\hline
\end{tabular}
\end{center}
\caption{ \label{tab:sum} Summary of inclusive branching ratio
results.}
\end{table}

These particles all have significant components of $s\bar{s}$. Our
results show that $\eta'$ and $\phi$ are relatively rare in $D^0$
and $D^+$ decay while the $\eta$ which has a lower mass and a
significant light quark component, is produced at a significantly
higher rate. The $\eta$, $\eta'$ and $\phi$ are all produced at
higher rates in $D_s$ decays than the corresponding rates from $D$
decays. The ratio of rates is given in Table~\ref{tab:sumrat}. The
$\phi$ yield is 15 times higher in $D_s^+$ decays than in $D$
decays.

\begin{table}[htb]
\begin{center}
\begin{tabular}{lccc}
\hline

Ratio     & $\eta$ & $\eta'$ & $\phi$  \\
\hline $D_s^+/D^0$ & ~~~$2.47\pm 0.34\pm 0.18$~~~ & ~~~$3.51\pm
0.80\pm 0.27$~~~ &~~~$15.3\pm 1.6\pm 0.8$~~~\\
$D_s^+/D^+$ &$3.73\pm 0.57\pm 0.27$ & $8.37\pm 2.23\pm 0.64$ &
~~~$15.6\pm1.9\pm 0.8$~~~\\
 \hline\hline
\end{tabular}
\end{center}
\caption{ \label{tab:sumrat} Ratios of $D_s^+$ to $D$ yields. Common
systematic errors have been eliminated.}
\end{table}

The large asymmetry in the yields of these particles between $D_s$
and the lighter $D$ mesons will permit further studies of  $B_s$
decays at the $\Upsilon$(5S), and will be most useful for separating
$B_s$ from $B$ decays at hadron colliders.

\section{Acknowledgments}
We gratefully acknowledge the effort of the CESR staff in providing
us with excellent luminosity and running conditions.
D.~Cronin-Hennessy and A.~Ryd thank the A.P.~Sloan Foundation. This
work was supported by the National Science Foundation, the U.S.
Department of Energy, and the Natural Sciences and Engineering
Research Council of Canada.

\end{document}